\documentclass[aps,pra,reprint,superscriptaddress]{revtex4-2}
\usepackage{amsmath}
\usepackage{graphicx}
\usepackage{amsfonts}
\usepackage{physics}
\usepackage{amssymb}
\usepackage{xcolor}

\begin{document}

\title{Predicting Multipartite Entanglement in Quantum Circuits using Transformer}

\author{Darell Timothy Tarigan}
\affiliation{Department of Physics, Faculty of Science and Mathematics, Bandung Institute of Technology, Bandung 40132, Jawa Barat, Indonesia }
\author{Fadhil Fatih Shiddiq}
\affiliation{Department of Physics, Faculty of Science and Mathematics, Bandung Institute of Technology, Bandung 40132, Jawa Barat, Indonesia }
\author{Hadyan Luthfan Prihadi}
\affiliation{Research Center for Quantum Physics, National Research and Innovation Agency (BRIN), South Tangerang 15314, Indonesia.}
\author{Donny Dwiputra}
\affiliation{Research Center for Quantum Physics, National Research and Innovation Agency (BRIN), South Tangerang 15314, Indonesia.}
\author{Jusak S. Kosasih}
\affiliation{Theoretical High Energy Physics Group, Department of Physics, FMIPA, Institut Teknologi Bandung, Jl. Ganesha 10 Bandung, Indonesia. }
\author{Yanoar P. Sarwono}
\email{yano001@brin.go.id}
\affiliation{Research Center for Quantum Physics, National Research and Innovation Agency (BRIN), South Tangerang 15314, Indonesia.}
\author{Freddy Permana Zen}
\email{fpzen@itb.ac.id}
\affiliation{Theoretical High Energy Physics Group, Department of Physics, FMIPA, Institut Teknologi Bandung, Jl. Ganesha 10 Bandung, Indonesia. }
\affiliation{Indonesia Center for Theoretical and Mathematical Physics (ICTMP), Institut Teknologi Bandung, Jl. Ganesha 10 Bandung, 40132, Indonesia. }

\author{Rui-Qin Zhang}
\email{aprqz@cityu.edu.hk}
\affiliation{Department of Physics, City University of Hong Kong, Hong Kong SAR, China}

\begin{abstract}
Multipartite entanglement is an important property of a parameterized quantum circuit (PQC), especially for near-term hybrid quantum-classical algorithms. It characterizes how the PQC can effectively generate highly entangled states, often referred to as its entangling capability. However, measuring it is still computationally difficult as conventional Monte Carlo sampling scales unfavorably with system size. To address this challenge, we introduce a graph-based transformer surrogate that predicts both the first-order Meyer--Wallach measure ($Q_1$) and the second-order Scott measure ($Q_2$), which resolves entanglement structures indistinguishable under $Q_1$ alone. Our central contribution is the qubit-interconnected graph (QIG) encoding for transformer, in which each node represents a qubit and a weighted adjacency records entangling-gate multiplicity on each bond. Fused with a gate-level DAG encoder, this yields the QIG-Fusion model. Trained and evaluated on $50{,}000$ circuits spanning 4- to 8-qubit systems over a ten-seed protocol, QIG-Fusion attains RMSE as low as $0.037$ ($Q_2$) and $0.038$ ($Q_1$), with Spearman rank correlation up to $0.95$. This contribution significantly reduces the computational cost of Quantum Architecture Search (QAS), enabling efficient multipartite entanglement estimation for large-scale PQCs and facilitating the design of more expressive quantum circuits.
\end{abstract}

\maketitle

\section{Introduction}
In the absence of fault-tolerant quantum computers, the advantages of quantum computing depend on hybrid quantum-classical algorithms~\cite{Preskill2018}. These algorithms deploy classical optimization to minimize specific objective functions evaluated in quantum processors or simulations, from estimating the ground-state energy of molecular systems to solving complex optimization problems~\cite{Peruzzo2014, Kandala2017, Cerezo2021,Iryanti2025,Darmawan2025,Ramadhan2026a,Ramadhan2026b,Hakim2026}. A crucial step in the algorithm involves choosing a proper parameterized quantum circuit (PQC) ansatz, which contains the optimized parameter. The choice of a proper ansatz affects the computational load and accuracy of the algorithm~\cite{Cerezo2021}. Several quantities which are devoted to study quantitative measures for characterizing ansatz quality have been investigated in PQCs, such as expressibility, trainability, and others~\cite{Sim2019, Holmes2022,Ramadhan2026a}. Another important property is the ability of an ansatz to generate quantum correlation, which is commonly quantified by entangling capability \cite{Sim2019}. Since quantum entanglement is a fundamental and genuine quantum resource in quantum computers~\cite{Nielsen2010}, understanding how it is generated within PQCs is crucial for evaluating ansatz performance.\\
\indent In PQCs, entanglement is generated primarily through two-qubit gates. In general, it acts as an important safeguard against classical tractability; entanglement is widely regarded as an important resource for generating non-classical correlations that can contribute to quantum computational advantage~\cite{Jozsa2002} and is a necessary resource in quantum computational speed-ups~\cite{Vidal2003}. Furthermore, this is a mathematically necessary condition for making circuits more expressive~\cite{Sim2019}. However, too much entanglement could lead to undesirable circuits with barren plateaus, which are associated with a global distribution of entanglement \cite{Kim2022,Kim2023}. Therefore, the ability to quantify entanglement in PQCs is important for determining a proper ansatz for specific variational tasks.\\
\indent One of the most widely used entanglement measures is the entanglement entropy and quantum mutual information~\cite{Bennett1995, Groisman2004}, which are essentially limited to bipartite correlations~\cite{Meyer2002, Brennen2003}. Other measures, such as Concurrence, are defined for 2-qubit systems but lack a practical generalization to multipartite regime~\cite{Wootters1998, Mintert2005}. However, characterizing how entanglement is distributed across the entire system generally requires information beyond bipartite correlations. In many-body quantum systems and PQCs, entanglement is often distributed among many qubits simultaneously. Capturing such global correlations requires multipartite entanglement measures.\\
\indent Among the available multipartite measures, the Mayer-Wallach and Scott \cite{Scott2004} measures are particularly attractive because of their ability to characterize entangling capability of PQCs \cite{Sim2019}. In this work, we consider the second-order Scott measure $Q_2$ alongside the Meyer-Wallach measure, which is a first-order Scott measure $Q_1$. The characterization of randomness and entangling capability in PQCs has recently been extended by \cite{Correr2025}, demonstrated that higher-order entanglement measures, such as the Scott order-2 metric ($Q_2$), can capture subsystem correlations that are not detected by the standard Meyer--Wallach measure. This highlights the need for a more comprehensive characterization of PQCs through multipartite entanglement. Consequently, entangling capability should be evaluated alongside other important properties, including expressibility and trainability, rather than arbitrarily maximized.\\
\indent Computing the entangling capability of parameterized quantum circuits (PQCs) is computationally expensive because it requires Monte Carlo sampling over thousands of random parameter realizations to estimate multipartite entanglement measures \cite{Sim2019, Aktar2023}. This cost becomes a major bottleneck for Quantum Architecture Search (QAS), where thousands of candidate circuits must be evaluated during the search process \cite{Martyniuk2024}. Surrogate models therefore offer a promising alternative by predicting circuit properties directly from structural representations. Recent work has demonstrated the potential of graph-based Transformers for predicting expressibility-related metrics \cite{Zhang2025}; however, accurately predicting entanglement additionally requires circuit representations that preserve the topological structure and parallelism of PQCs, which are often lost in conventional sequence- and matrix-based encodings \cite{Martyniuk2024}. To address these challenges, we propose \textbf{QIG-Fusion}, a Transformer-based surrogate model that combines a qubit-interconnected graph with a gate-level directed acyclic graph (DAG) encoding to predict multipartite entanglement, enabling efficient pre-screening of candidate circuits for scalable QAS.\\
\indent To the best of our knowledge, Transformer-based surrogate models have not previously been developed for predicting the entangling capability of PQCs. The closest related work is by Zhang \textit{et al.} \cite{Zhang2025b}, who employed an LSTM with a gate-to-tensor encoding to predict the Meyer--Wallach entangling capability together with trainability for 8-qubit circuits, primarily to investigate the entangling capability and trainability. In contrast, our work differs in three key aspects, and these are the main contributions of this work: (i) we employ a graph Transformer with the proposed qubit-interconnected graph encoding; (ii) we predict higher-order multipartite entanglement through the second-order Scott measure ($Q_2$) in addition to the first-order Meyer--Wallach measure ($Q_1$); and (iii) we evaluate the model across 4--8 qubit circuits using a ten-seed experimental protocol.

\section{Theory}

A PQC operates on a quantum register consisting of $n$ qubits~\cite{Sim2019}. The system is initialized in the computational basis state, denoted mathematically as $\ket{0}^{\otimes n}$~\cite{Cerezo2021}. The evolution of the quantum state is driven by a sequence of quantum logic gates, which can be collectively represented by a unitary operator $U(\vec{\theta})$~\cite{Cerezo2021, Grant2019}. $U(\vec{\theta})$ denotes the sequence of quantum gates, encompassing both parameterized single-qubit rotation gates, such as $U_3$, and two-qubit entangling gates, such as $CZ$, while $\vec{\theta}$ represents the vector of trainable gate parameters~\cite{Sim2019, Zhang2025b}. The output state can be written as,
\begin{equation}
    \ket{\psi_\theta} = U(\theta)\ket{0}^{\otimes n}.
\end{equation}

To comprehensively evaluate the entangling capability of PQC, choosing an appropriate mathematical metric is crucial~\cite{Sim2019, Hubregtsen2020}. Conventional metrics such as von Neumann entropy and concurrence are fundamental in quantum information theory~\cite{Nielsen2010}, but are essentially limited to evaluating bipartite entanglement, strictly measuring quantum correlations between two isolated subsystems or specific pairs of qubits~\cite{Bennett1995, Wootters1998}. In the context of multi-qubit PQC architectures, entanglement is rarely confined to localized pairs, but rather spreads across registers as a complex phenomenon throughout the system~\cite{Sang2023, Meyer2002}. Consequently, bipartite metrics are inadequate for this study because they fail to capture the holistic and global entanglement topology generated by an ansatz~\cite{Meyer2002, Scott2004}. Bipartite entanglement describes non-local quantum correlations between two different subsystems~\cite{Bennett1995, Horodecki2009}. This is a well-understood form of entanglement, and measures such as concurrence and von Neumann entropy of the reduced density matrix can be used to fully characterize it, but fail to fully characterize the complex structure of multipartite entanglement. On the other hand, multipartite entanglement involves correlations shared by three or more subsystems that cannot be factored into separate bipartite states or single-qubit states~\cite{Coffman2000, Dur2000}. Examples are the GHZ state ($\ket{000} + \ket{111}$) and the W state~\cite{Dur2000}. Mathematically, it is difficult to quantify this because entanglement can be distributed in various ways across the network~\cite{Horodecki2009, Dur2000}. Metrics such as the Meyer-Wallach measure or global entanglement are needed to quantify it.

To overcome this limitation and rigorously quantify the global entangling strength of a circuit, we rigorously employ a multipartite entanglement measure~\cite{Scott2004}. Specifically, we use a general measure introduced by Scott. Scott's framework resolves the mathematical complexity of multipartite entanglement by proposing a systemic evaluation across all possible subsystem subdivisions~\cite{Scott2004, Haddadi2018}. Instead of relying on isolated pairs, the $m$-th order Scott measure ($Q_m$) evaluates the average entanglement across all possible subsystems of size $m$. Within this framework, the well-known Meyer-Wallach measure ($Q_1$) naturally emerges as a first-order ($m = 1$) formulation of the Scott measure~\cite{Brennen2003, Scott2004}. The Meyer-Wallach metric quantifies global entanglement by evaluating the linear entropy of the reduced density matrix for each individual qubit after tracing out the rest of the system, then averaging these values. By leveraging the Meyer-Wallach measure and the Scott measure, this study ensures a robust and comprehensive characterization of global entanglement by capturing single-qubit to multi-qubit interactions as well as higher-order pair correlations that would be completely ignored by simpler metrics~\cite{Sim2019, Scott2004}.

The capacity of PQC to generate entanglement is intrinsically linked to its average ability to generate entangled states, a property typically quantified through state entanglement metrics~\cite{Sim2019}. As a comprehensive global metric for multi-particle entanglement in pure states, the Meyer-Wallach measure has been widely adopted in various quantum information domains~\cite{Meyer2002, Brennen2003}, due to its robust characterization of entanglement properties. Consequently, it serves as an optimal tool for assessing the entangling capability of PQC by analyzing the output states it produces~\cite{Sim2019}. In PQC, the Meyer-Wallach (MW) entangling capability measures the average ability of a quantum circuit to generate multiparticle entanglement across the entire quantum system~\cite{Sim2019, Meyer2002}. For an $n$-qubit system, the MW entanglement, denoted as $E_{MW}$, is mathematically formulated as~\cite{Meyer2002, Brennen2003}
\begin{equation}
    E_{\text{MW}}(\ket{\psi}) \equiv \frac{4}{n} \sum_{i=1}^{n} D\!\left(\Gamma_i^{(0)}\ket{\psi}, \Gamma_i^{(1)}\ket{\psi}\right),
\end{equation}
where $\Gamma_i^{(b)}$ represents a linear mapping applied to the computational basis, $b \in \{0,1\}$. This mapping is defined by
\begin{equation}
    \Gamma_i^{(b)}\ket{b_1 \ldots b_n} = \delta_{b,b_i}\ket{b_1 \cdots \ominus b_i \ldots b_n},
\end{equation}
where the $\ominus$ symbol indicates the operation of tracing out the $i$-th qubit. The operator $D$ indicates the general distance, which is calculated through
\begin{equation}
    D(\ket{u}, \ket{v}) = \sum_{i,j} \frac{1}{2}\left|u_i v_j - u_j v_i\right|^2,
\end{equation}
assuming the representation of the quantum state is $\ket{u} = \sum_i u_i\ket{i}$ and $\ket{v} = \sum_i v_i\ket{i}$. The Meyer-Wallach entanglement capability ($Q_1$) of a given PQC is determined by calculating the average MW entanglement ($E_{\text{MW}}$) across the entire set of output states, randomly sampled from the circuit parameter space $\theta_i$. This relationship is expressed as
\begin{equation}
    Q_1 \equiv \overline{E_{\text{MW}}} \equiv \frac{1}{|S|} \sum_{\theta_i \in S} E_{\text{MW}}(\ket{\psi_{\theta_i}}).
\end{equation}
$S = \{\theta_i\}$ denotes the set of parameter vectors sampled from the PQC parameter space, and $|S|$ denotes the total number of vectors sampled. To ensure statistical convergence of the calculated $Q_1$, sufficient sampling density in the parameter space is required, which needs to be evaluated to achieve stable and representative estimates~\cite{Sim2019, Aktar2023}.

While the Meyer-Wallach measure is very effective, it relies on the average entanglement of single qubits~\cite{Meyer2002, Brennen2003}. To capture higher-order multipartite entanglement that might be overlooked by the Meyer-Wallach measure, we use a general measure introduced by Scott~\cite{Scott2004}. Measuring entanglement across a massive system consisting of many qubits is considered mathematically complex~\cite{Scott2004, Horodecki2009}. Scott's goal was to show that complex mathematics is not strictly necessary to measure multipartite entanglement; instead, one can average the bipartite entanglement over all possible subdivisions of the system. Scott generalizes the Meyer-Wallach measure by extending the notion of global entanglement from single-qubit reductions to arbitrary $m$-qubit subsystems. Instead of simply considering the average purity of single-qubit reduced states, the $m$-order Scott measure evaluates the average purity over all possible $m$-qubit subsystems, thereby capturing the higher-order correlation structure in multipartite quantum systems~\cite{Scott2004, Aaronson2004}. Formally, the measure is defined as
\begin{equation}
    Q_m(\psi) \equiv \frac{D^m}{D^m - 1}\left(1 - \frac{m!\,(n-m)!}{n!} \sum_{|S|=m} \tr(\rho_S^2)\right),
\end{equation}
where $D$ denotes the local Hilbert space dimension of each subsystem ($D = 2$ for a qubit system), $n$ is the total number of qubits, and $S$ represents an $m$-qubit subsystem. The second-order Scott measure is then written as
\begin{equation}
    E_2(\ket{\psi_{\theta_i}}) \equiv \frac{4}{3}\left(1 - \frac{2}{n(n-1)} \sum_{|S|=2} \tr(\rho_S^2)\right),
\end{equation}
and
\begin{equation}
    Q_2 \equiv \frac{1}{|S|} \sum_{\theta_i \in S} E_2(\ket{\psi_{\theta_i}}).
\end{equation}

Different orders $m$ of the Scott measure probe different facets of multipartite entanglement, and a first-order measure alone can be degenerate. A canonical illustration is provided by two six-qubit states: the GHZ state $\ket{\mathrm{GHZ}_6}=\left(\ket{0}^{\otimes 6}+\ket{1}^{\otimes 6}\right)/\sqrt{2}$ and the triple-EPR state $\ket{\mathrm{EPR}_6}=\ket{\Phi^+}^{\otimes 3}$,  $\ket{\Phi^+}=\left(\ket{00}+\ket{11}\right)/\sqrt{2}$. Under the first-order Meyer--Wallach measure, these states are indistinguishable, with $Q_1(\ket{\mathrm{GHZ}_6})=Q_1(\ket{\mathrm{EPR}_6})=1$, even though their entanglement structures are fundamentally different: $\ket{\mathrm{GHZ}_6}$ exhibits genuinely global multipartite correlations, whereas $\ket{\mathrm{EPR}_6}$ consists of three independent Bell pairs and is therefore biseparable . Choosing $m=2$ resolves this degeneracy, yielding $Q_2(\ket{\mathrm{GHZ}_6})=2/3$ and $Q_2(\ket{\mathrm{EPR}_6})=4/5$, which clearly distinguishes the two states. Because quantum architecture search is concerned with the  of entanglement that a circuit can generate, not merely its average bipartite magnitude, predicting $Q_2$ alongside $Q_1$ provides a richer characterization of a circuit's entangling capability and enables us to evaluate whether our findings remain consistent across both measures\cite{Correr2025}.

An important constraint of the Scott measure is that it requires $n \geq m + 1$ qubits, which is nontrivial when $n = m$, the subsystem encompasses all pure states, resulting in $\tr(\rho_S^2) = 1$ and consequently $Q_m = 0$~\cite{Scott2004, Haddadi2018}. Furthermore, for $n = m + 1$, only one qubit is traced out, severely limiting the achievable range of $Q_m$. For statistically meaningful claims about higher-order multipartite entanglement, we therefore require $n \geq m + 2$, which ensures at least two qubits are traced out and the reduced state can exhibit non-trivial mixing. $Q_4$ is identically zero for 4-qubit circuits. The Scott measure obeys fundamental geometric ordering on small-qubit systems. For a pure $n$-qubit state with $n < 2m$, the maximum attainable $Q_m$ is bounded by the geometric structure of the bipartition $|S| = m$ versus $|S| = n - m$: a reduced state on $S$ cannot exceed rank $\min(D^m, D^{n-m})$~\cite{Du2022}, which leads to the bound
\begin{equation}
    Q_m \leq \frac{D^m}{D^m - 1}\left(1 - \frac{1}{\min(D^m, D^{n-m})}\right).
\end{equation}
Given that there are 4, 5, and 6 qubits, the focus in this work is on Scott order 2 ($Q_2$) and Scott order 1 ($Q_1$) or Meyer-Wallach. The Haar mean of the Scott measure, denoted by
\begin{equation}
    \langle Q_m\rangle_{\mathrm{Haar}} = \frac{2^m}{2^m - 1}\left(1 - \frac{2^m + 2^{n-m}}{2^n + 1}\right),
    \label{eq:haar}
\end{equation}
represents the expected entanglement of randomly distributed quantum states and is commonly used as a benchmark for evaluating the entanglement capability of quantum circuits~\cite{Sim2019, Scott2004}. $Q_m$ value close to the Haar mean indicates that the quantum circuit is capable of strong, near-maximal, multipartite entanglement~\cite{Sim2019, Scott2004}.

The transformer model is adopted in this study to map the complex and non-linear relationship between circuit architecture and the resulting entangling capability~\cite{Zhang2025, Vaswani2017}. The transformer model is used because it has the robust ability to capture complex relationships and long-range dependencies in quantum circuit representations of varying lengths, making it suitable for studying the relationship between PQC structure and its expressibility. Similar to the challenges faced in natural language processing~\cite{Vaswani2017}, PQC exhibits highly variable lengths in terms of the number of qubits and total gate operations. The transformer architecture is naturally well-suited to processing such variable-length inputs without requiring rigid dimensionality constraints~\cite{Vaswani2017}.

The advantage of our approach lies in its ability to overcome some of the structural limitations of sequential-based surrogate models, such as Long Short-Term Memory (LSTM) and Recurrent Neural Networks (RNN)~\cite{Vaswani2017, Hochreiter1997}. These approaches typically represent quantum circuits as strict temporal sequences, where each time step represents only one quantum gate, and standard RNNs including LSTMs are limited in their capability to handle tasks involving very long sequences because computation of features for different parts cannot occur in parallel~\cite{Hochreiter1997, Bradbury2016}. This serialization process obscures the intrinsic concepts of circuit depth and gate parallelism inherent in quantum hardware~\cite{Aktar2023, Cross2017}. For example, our 6-qubit dataset has circuits with up to 39 gates but only a physical depth of 23 layers (Table~\ref{tab:dataset}), meaning a naive sequential encoding would increase the effective sequence length by approximately 70\%~\cite{Zhang2025}. While manageable for small circuits, this inflation becomes increasingly problematic in deeper QAS workflows involving hundreds of gates~\cite{Zhang2025, Du2022}. Furthermore, operations on the same circuit layer acting on different subsets of qubits are causally independent, whereas the sequential model imposes an artificial order between them~\cite{Cross2017}. As a result, sequential encoding can fail to faithfully preserve the underlying causal structure and topology of entanglement propagation in local quantum systems~\cite{Sang2023, Cross2017}.

Our  framework addresses this critical computational hurdle by synergizing the transformer architecture with directed acyclic graph (DAG) encoding~\cite{Zhang2025, Cross2017}. Unlike one-dimensional sequential arrays, the DAG representation naturally respects the total number of gates and the actual physical circuit depth. In a DAG, gates operating in parallel are accurately represented as independent nodes within the same topological layer~\cite{He2016}, essentially preventing artificial elongation of the input sequence. Therefore, directed acyclic graph (DAG)-based encoding was chosen because it better preserves global topological information and dependency relationships between gates in quantum circuits than image-based encoding methods~\cite{Aktar2023}, which struggle to accurately represent topological relationships and non-adjacent two-qubit gates, or sequence, which removes parallelism information and imposes an artificial order between gates, or binary, which lacks the flexibility to efficiently represent complex circuit structures and continuous parameters~\cite{Du2022}.

The self-attention mechanism of the multi-head transformer is uniquely equipped to process these graphical structures, allowing the model to jointly attend to information from different representation subspaces at different positions~\cite{Vaswani2017}. Instead of reading the circuit step by step, the self-attention mechanism processes the entire graph simultaneously with a global receptive field. Because parallel gates at the same depth are represented as independent nodes while the directed adjacency records their downstream convergence, global attention can relate gates that jointly feed a later entangling operation without imposing an artificial sequential order.\\

\indent Furthermore, we empirically demonstrate the structural robustness of this DAG-transformer paradigm by evaluating it on higher-order pairwise multipartite entanglement, specifically, $Q_2$ together with conventional $Q_1$. Ultimately, this establishes DAG-encoded transformers as a robust and effective framework for machine learning algorithms to read and evaluate the physical properties of quantum circuits.\\
\begin{figure*}[!htbp]
    \centering
    \includegraphics[width=\textwidth]{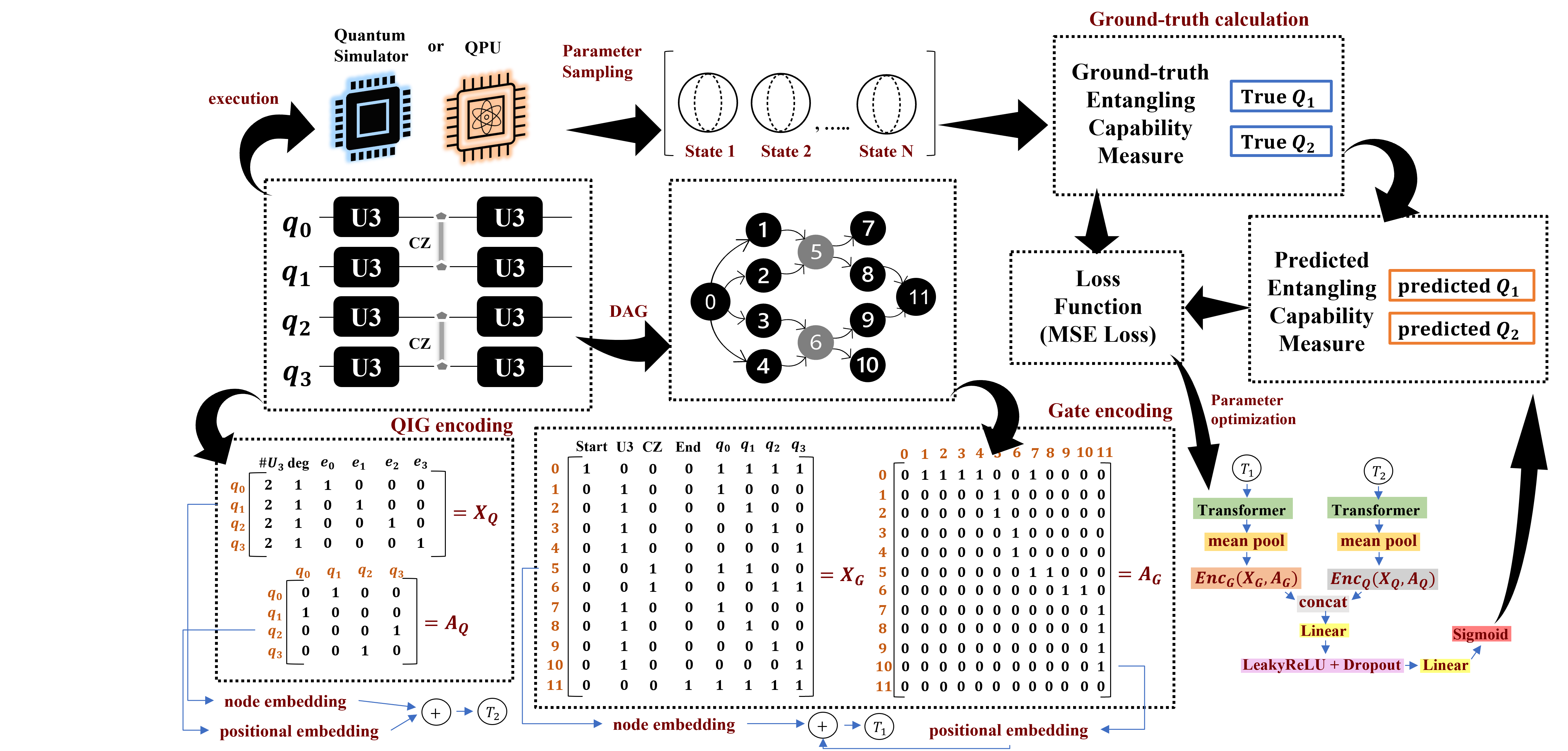}
    \caption{This pipeline consists of three stages: (1) gate-level and QIG encoding of PQC; (2) ground truth entanglement evaluation for $Q_1$ and $Q_2$ using quantum simulation; and (3) transformer-based learning to predict the entanglement measure.}
    \label{fig:pipeline}
\end{figure*}
However gate-level encodings represent a circuit as a DAG of gates, so the representation size grows with circuit length and the same physical qubit appears in many nodes. Because our prediction targets $Q_1$ and $Q_2$ are ensemble averages over the circuit's parameters, they depend on the circuit only through its entangling structure: which qubits interact, how often, and how much local rotation each carries. The QIG encoding is designed to expose exactly this structure.  An overview of the framework and examples of the two graph encodings are illustrated in Figure~\ref{fig:pipeline}  through the feature matrices $X_Q$, $X_G$ and the adjacency matrices $A_Q,A_G$. Let a parameterized quantum circuit on $n$ qubits be an ordered list of gates $\mathcal{G} = (g_1,\dots,g_G)$, where each $g$ is either a single-qubit rotation $U_3$ acting on qubit $a(g)$, or a two-qubit entangler $\mathrm{CZ}$ acting on the ordered pair $(a(g),b(g))$,  $\mathcal{U} = \{g : g \text{ is } U_3\}$ and $\mathcal{C} = \{g : g \text{ is } \mathrm{CZ}\}$. The qubit-interconnected graph is the weighted undirected graph with $n$ nodes, one per qubit. Its weighted adjacency matrix $A_Q \in\mathbb{Z}_{\ge 0}^{\,n \times n}$ counts entangler multiplicity on each bond:
\begin{equation}
\label{eq:qig-adj}
(A_Q)_{ij} \;=\; \bigl|\{\, g \in \mathcal{C} \;:\; \{a(g),b(g)\} = \{i,j\} \,\}\bigr|,
\end{equation}
where $(A_Q)_{ii}=0$ and $A_Q=A_Q^{\top}$. The node-feature matrix $X_Q \in \mathbb{R}^{n \times (n+2)}$ assigns to each qubit $i$, its local rotation count, entangling degree, and a positional one-hot
\begin{equation}
\label{eq:qig-feat}
(X_Q)_{i,:} =
\bigl[r_i \mid \deg_i \mid \mathbf{e}_i^{\top}\bigr],
\end{equation}
\begin{equation}
\label{eq:qig-stat}
r_i =
\left|\left\{\, g\in\mathcal{U}:a(g)=i\,\right\}\right|,
\qquad
\deg_i =
\sum_{j=1}^{n} (A_Q)_{ij}.
\end{equation}
where $\mathbf{e}_i \in \{0,1\}^n$ is the $i$-th standard basis vector. Thus $r_i$ is the number of $U_3$ gates on qubit $i$, $\deg_i$ its weighted degree, and $\mathbf{e}_i$ breaks the permutation symmetry among otherwise identical nodes. The encoding has three properties that motivate its use seperti  the graph has exactly $n$ nodes irrespective of circuit depth or gate count. $X_Q$ and $A_Q$ depend only on gate placement, never on  rotation angles. The encoding is invariant under any permutation of the gate list that preserves per-qubit gate counts and bond multiplicities. This is a deliberate abstraction that discards temporal ordering, which the gate-level encoder retains.\\
\indent For comparison and as the second fusion branch, we retain a gate-level encoding which each of the $G$ gate is a node (plus start/end nodes). Its feature matrix $\mathbf{X}_G \in \mathbb{R}^{(G+2)\times(N_{gt}+n)}$ concatenates a gate-type one-hot ($N_{gt}$ categories) with a qubit one-hot, and its binary adjacency $\mathbf{A}_G \in \{0,1\}^{(G+2)\times(G+2)}$ encodes temporal dependency between gates. Unlike the QIG, the node count grows with circuit length and is padded to a fixed maximum. Both branches use the same encoder. Given a node-feature matrix  $F \in \mathbb{R}^{N \times d_{in}}$ and adjacency $M \in \mathbb{R}^{N \times N}$, the input embedding for node $i$ is
\begin{equation}
\label{eq:embed}
h_i^{(0)} \;=\; W_x\, F_{i,:}^{\top} \;+\; W_A\, M_{i,:}^{\top},
\end{equation}
\begin{equation}
\qquad W_x \in \mathbb{R}^{d \times d_{in}}, \qquad W_A \in \mathbb{R}^{d \times N},
\end{equation}
where d is the model width. The embeddings pass through $L$  transformer encoder layers with $n_h$ heads, followed by mean-pooling over nodes:
\begin{equation}
\label{eq:pool}
\text{Enc}(F,M) \;=\; \frac{1}{N} \sum_{i=1}^{N} h_i^{(L)} \;\in\; \mathbb{R}^{d}.
\end{equation}
Mean-pooling maps any node count $N$ to a fixed $d$-dimensional summary, so the $n$-node QIG and the $(G+2)$-node DAG produce vectors of identical size. The gate-level graph $(\mathbf{X}_G, \mathbf{A}_G)$ and the qubit-interconnected graph $(\mathbf{X}_Q, \mathbf{A}_Q)$ are processed by two architecturally identical but independently parameterized encoders, $\mathrm{Enc}_G$ and $\mathrm{Enc}_Q$; their pooled embeddings are concatenated and mapped to a scalar prediction:
\begin{equation}
\label{eq:fusion}
\begin{aligned}
\hat{Q}
&= \sigma\!\left(
\mathrm{MLP}\!\left(
\left[
\mathrm{Enc}_G(X_G,A_G)
\;\Vert\;
\mathrm{Enc}_Q(X_Q,A_Q)
\right]
\right)\right),
\end{aligned}
\end{equation}
$\bigl[\,\cdot \;\Vert\; \cdot\,\bigr] \in \mathbb{R}^{2d}$ and $\sigma$ is the logistic function, matching the range of the normalized targets
$Q_1, Q_2 \in [0,1]$. The MLP is a two-layer head
$\mathbb{R}^{2d} \to \mathbb{R}^{d} \to \mathbb{R}$ with LeakyReLU and dropout.\\
\indent The two branches are complementary by construction: the QIG captures aggregate interaction structure but is blind to gate order, while the DAG captures temporal dependency and represents interaction structure only implicitly. Their combination is not merely a consequence of increased model capacity; parameter-matched controls (Figure~\ref{fig:fiveencoder}) show that a dual gate-level encoder of equal size confers no benefit, while the addition of the QIG branch does. We adopt $L=2$, $n_h=2$, and $d=32$ per encoder, based on the ablation study in Figure~\ref{fig:ablation}; the fusion model contains $\approx 5.5\times10^5$ parameters.\\
\indent Circuit structure generation is performed using a stochastic algorithm, a gatewise pipeline~\cite{Zhang2025}, to ensure diversity in the architecture search space. This method works constructively by sequentially constructing a gate-by-gate circuit, starting with an empty qubit initialization. At each gate addition step, the algorithm performs probabilistic sampling to determine the gate type ($U_3$ or $CZ$) and its application location on the qubit. This system does not construct gates by deterministic selection, but rather samples from a probability distribution. The gate type is determined by sampling from a probability vector that follows a Gaussian distribution $\mathcal{N}(0, 1.35)$~\cite{Zhang2025}. This allows for a wide range of parameter-to-entanglement ratios between circuits. The use of a Gaussian distribution in the gate selection probability vector aims to prevent bias toward specific structural patterns, allowing the transformer model to learn a more general circuit representation. Once a gate type is selected, the algorithm must determine which qubit pair will perform this operation. The gate placement at the qubit index is determined using a normal distribution $\mathcal{N}(0, 1)$~\cite{Zhang2025}, which prevents gate overcrowding on a particular qubit and promotes more even circuit depth~\cite{Aktar2023, Zhang2025}. This dataset is purposefully designed to cover a wide range of circuit complexities and topologies, allowing the transformer model to effectively generalize to a wide range of PQC structures while avoiding overfitting to specific classes of quantum circuits. Furthermore, nearest-neighbor qubit connectivity constraints are used, where two-qubit $CZ$ gates are only applied to adjacent qubits in a linear topology, to more realistically represent the physical limitations of practical quantum hardware~\cite{Sim2019, Hubregtsen2020}.

\begin{table}[!htbp]
    \centering
    \caption{Circuit Dataset Information with Number of Circuits: 10,000 per qubit count}
    \label{tab:dataset}
    \begin{ruledtabular}
    \begin{tabular}{ccccc}
        Number of & Total Number & Circuit & Number of\\
        Qubits ($N_q$) & of Gates (Range) & Depth & Param. $U_3$ Gates \\
        \colrule
        4 Qubit & 10 -- 29 & 4 -- 22 & 4 -- 20 \\
        5 Qubit & 15 -- 34 & 5 -- 22 & 5 -- 24 \\
        6 Qubit & 20 -- 39 & 6 -- 23 & 6 -- 28 \\
        7 Qubit & 25 -- 44 & 5 -- 27 & 7 -- 31 \\
        8 Qubit & 30 -- 49 & 6 -- 27 & 8 -- 35 \\
    \end{tabular}
    \end{ruledtabular}
\end{table}

To rigorously validate our choice of 10,000 circuits per qubit count, we performed a two-stage convergence analysis presented in Figures~\ref{fig:resampling} and~\ref{fig:rmse_size}. First, we performed a distribution stability analysis under resampling: for each metric $Q_m$ and each subsample size $N \in \{500, 1000, 3000, 5000\}$, we took $R = 100$ independent subsamples from the full dataset and measured the Wasserstein distance $W_2$ between the empirical distribution and the reference population~\cite{Villani2009, Panaretos2019}. Across all four metrics, $W_2$ decreased by about 78\% from $N = 500$ to $N = 5000$, with the $\pm 1\sigma$ band narrowing tightly around the population curve at $N \geq 3000$, confirming statistical convergence~\cite{Panaretos2019}. Second, we evaluated the predictive RMSE as a function of training set size, and found that the improvement beyond $N = 5000$ was marginal (below 1.5\%), justifying our choice of 8,000 training circuits per configuration also in the saturation regime~\cite{Goodfellow2016}.

To validate the performance of the transformer model in reproducing the physical characteristics of the circuit without full quantum simulation, four complementary evaluation metrics were used~\cite{Aktar2023, Sim2019}. The root mean square error (RMSE) 
\begin{equation}
    \mathrm{RMSE} = \sqrt{\frac{1}{M} \sum_{i=1}^{M} (\hat{y}_i - y_i)^2},
\end{equation}
measures the absolute deviation between the predicted value $\hat{y}_i$ and the actual value $y_i$~\cite{Chai2014}. Here, $M$ denotes the total number of evaluated samples and $R^2$ is the coefficient of determination, which measures the agreement between predicted values and observed data by measuring the proportion of variance in the target variable explained by the model~\cite{Chicco2021}. A high RMSE value indicates lower accuracy, while a lower value indicates higher numerical accuracy, expressed as~\cite{Chai2014, Hyndman2006}. The variable $\bar{y}$ represents the average of the actual values. A coefficient of determination value close to 1 indicates a strong agreement between predictions and observations, and vice versa. The $R^2$ is defined as~\cite{Chicco2021}
\begin{equation}
    R^2 = 1 - \frac{\sum_{i=1}^{M} (y_i - \hat{y}_i)^2}{\sum_{i=1}^{M} (y_i - \bar{y})^2}.
\end{equation}
In addition to per-point accuracy, the model's ability to preserve the relative order of entanglement measures is crucial for analytical optimization and architecture search~\cite{Zhang2025}. Therefore, we use two rank-based metrics~\cite{Spearman1904, Kendall1938}. Spearman's rank correlation coefficient defined as
\begin{equation}
    \rho = \frac{\sum_{i=1}^{M} (r(\hat{y}_i) - \overline{r_{\hat{y}}})(r(y_i) - \overline{r_y})}{\sqrt{\sum_{i=1}^{M} (r(\hat{y}_i) - \overline{r_{\hat{y}}})^2 \sum_{i=1}^{M} (r(y_i) - \overline{r_y})^2}},
\end{equation}
evaluates the monotonicity of the relationship between predicted and target values by comparing their respective ranks, $r(\cdot)$ denotes the ranking operator~\cite{Spearman1904}
Kendall's tau coefficient ($\tau$) is used as an additional robust rank-based metric to assess the consistency between predicted and target sequences~\cite{Kendall1938}. It is defined as the difference between the number of concordant pairs $n_c$ and discordant pairs $n_d$ among all possible $\binom{M}{2}$ pairs of observations, which can be written as
\begin{equation}
    \tau = \frac{n_c - n_d}{\frac{1}{2} M(M-1)}.
\end{equation}

\section{Result \& Discussion}

\begin{figure*}[!htbp]
    \centering
    \includegraphics[width=0.85\textwidth]{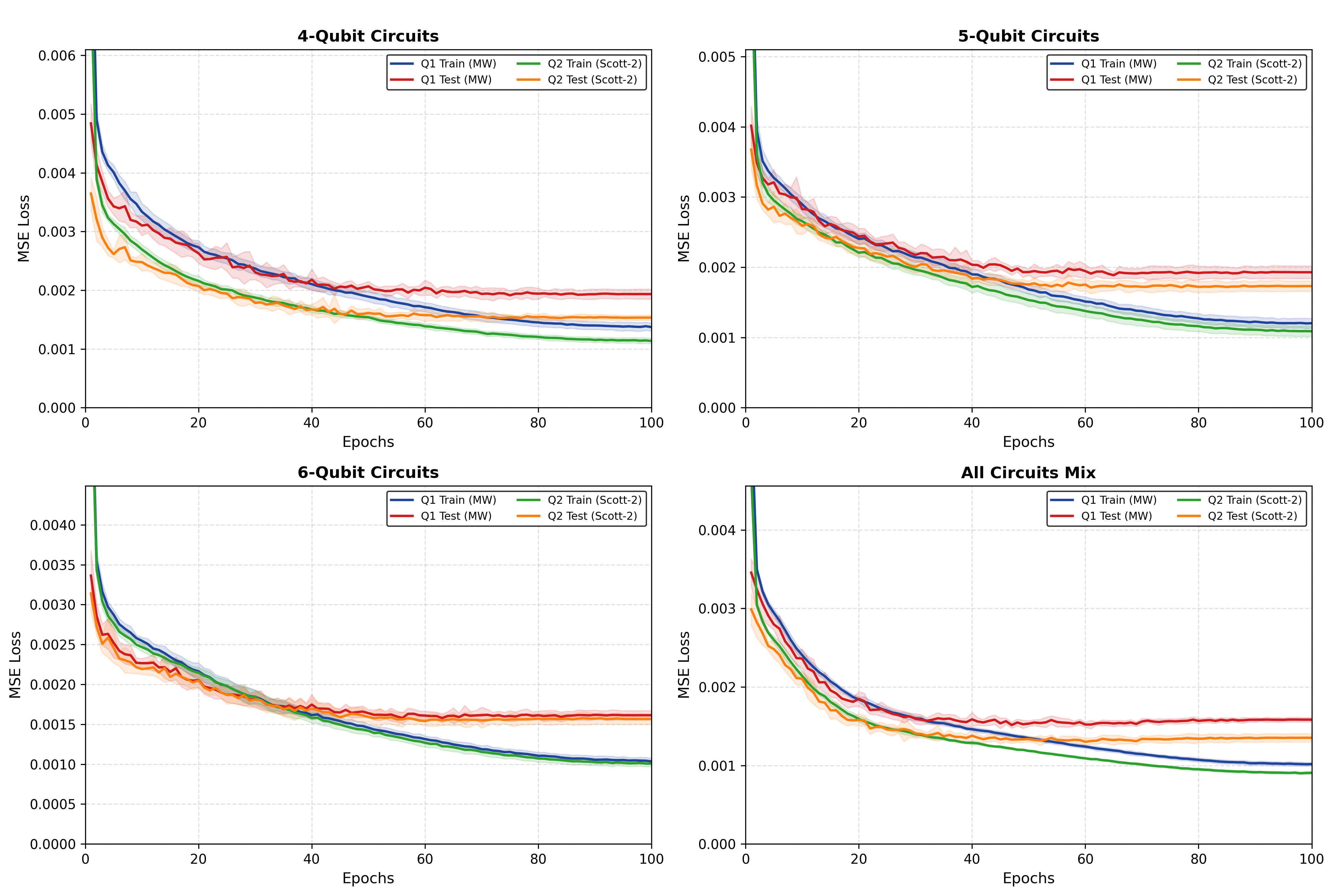}
    \caption{Training and testing loss convergence curves of the optimal transformer model $(2$-$2$-$32)$ over 100 epochs across different qubit scenarios for $Q_1$ \& $Q_2$.}
    \label{fig:loss}
\end{figure*}

To ensure full reproducibility and rigorous statistical reporting consistent, we adopt an audit-aligned training protocol comprising the following components: For per-qubit-count datasets ($N_q = 4, 5, 6$), each containing 10,000 circuits, we apply a fixed random seed ($\mathrm{seed} = 42$) to perform an $80{:}20$ train-test split. For the combined/Mix dataset of 30,000 circuits, we perform stratified splitting per qubit count to guarantee balanced $1{:}1{:}1$ representation across the train and test partitions, ensuring that observed performance on Mix datasets does not arise from overrepresentation of any single qubit count. For the 7- and 8-qubit datasets, each also containing 10,000 circuits, we follow the same train-test splitting protocol. However, these datasets use a different padding configuration to accommodate the larger graph size. Each model configuration is trained independently across 10 random seeds $\{0, 1000, 2000, \ldots, 9000\}$, with full reproducibility ensured via deterministic seeding of PyTorch, CUDA, NumPy, and Python random states. All reported metrics are aggregated as mean $\pm$ standard deviation across these 10 runs. Following best practices in machine-learning evaluation~\cite{Ferrer2024}, we intentionally use no validation set, training proceeds for a fixed 100 epochs, and the final epoch model is retained without any selection based on test loss. Test loss is monitored solely for diagnostic visualization (Figure~\ref{fig:loss}) and never enters the model selection loop. This protocol eliminates the optimistic bias that arises when test sets are implicitly used for model selection.

\begin{figure*}[!htbp]
    \centering
    \includegraphics[width=0.95\textwidth]{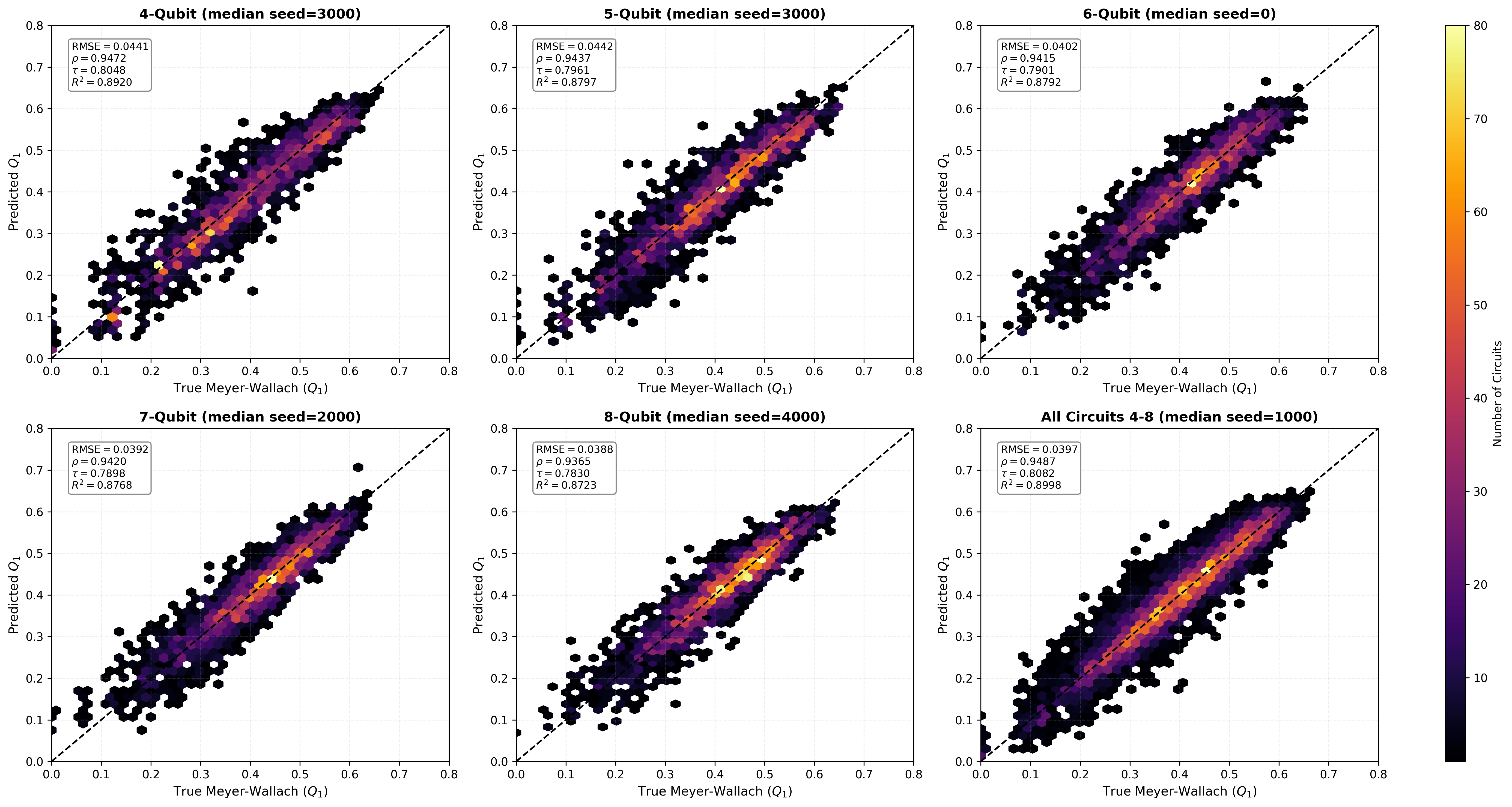}
    \caption{Correlation between the predicted and target Meyer-Wallach entanglement measures (Q1) on the test set, shown for the median-performing seed across 10 independent runs. Inset statistics correspond to the displayed median-RMSE seed; values reported in the main text are the mean ± standard deviation over 10 seeds. ; 4-, 5-, and 6- qubit circuits use padding (41 node, 10 feature dim), while 7- and 8- qubit circuits use padding (51 node, 12 feature dim).}
    \label{fig:mw_corr}
\end{figure*}
Figure~\ref{fig:mw_corr} compares the $Q_1$ values predicted by the QIG-Fusion model with the simulated ground truth; the panels show the median-RMSE seed, while the statistics quoted here are the mean $\pm$ standard deviation over ten independent seeds. The model attains RMSE of $0.0440 \pm 0.0010$, $0.0439 \pm 0.0010$, and $0.0402 \pm 0.0007$ for the 4-, 5-, and 6-qubit datasets, $0.0396 \pm 0.0007$ and $0.0385 \pm 0.0009$ for 7 and 8 qubits, and $0.0398 \pm 0.0003$ on the combined 4--8 dataset, with Spearman $\rho \approx 0.94$--$0.95$ and Kendall $\tau \approx 0.78$--$0.81$. The small standard deviations confirm stable performance across seeds, and the mean RMSE decreases monotonically as circuit size grows (from $0.0440$ at 4 qubits to $0.0385$ at 8 qubits), indicating that the graph-based representation remains effective at larger scales. This high rank fidelity is central to Quantum Architecture Search, where ordering candidate circuits by entanglement strength matters as much as an exact estimate~\cite{Du2022,Zhang2022}.

\begin{figure*}[!htbp]
    \centering
    \includegraphics[width=0.95\textwidth]{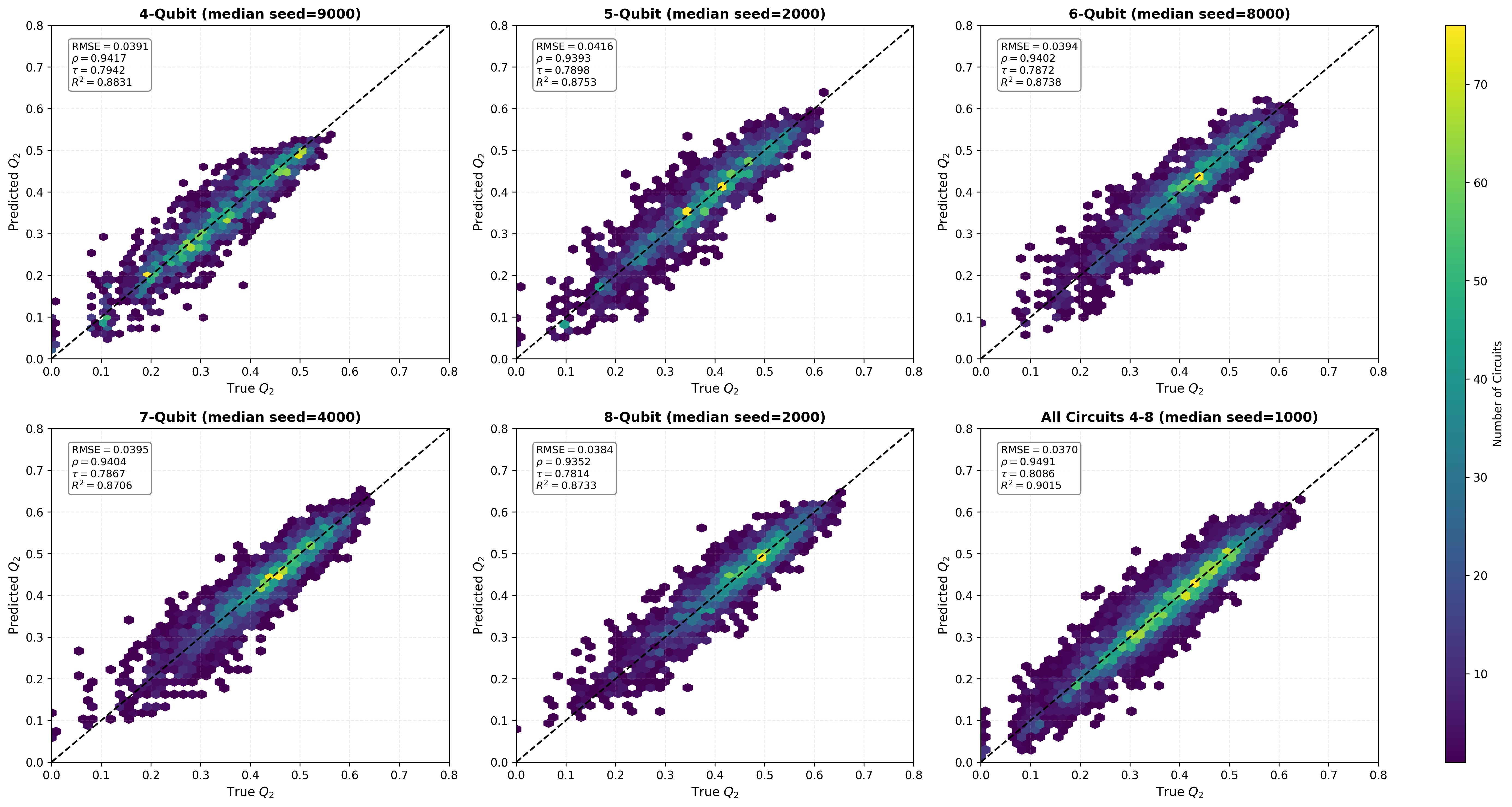}
    \caption{Correlation between the predicted second-order Scott entanglement measures (Q2) on the test set, shown for the median-performing seed across 10 independent runs. Inset statistics correspond to the displayed median-RMSE seed; values reported in the main text are the mean ± standard deviation over 10 seeds; 4-, 5-, and 6- qubit circuits use padding (41 node, 10 feature dim), while 7- and 8- qubit circuits use padding (51 node, 12 feature dim).}
    \label{fig:scott_corr}
\end{figure*}

\begin{figure}[!t]
    \centering
    \includegraphics[width=\columnwidth]{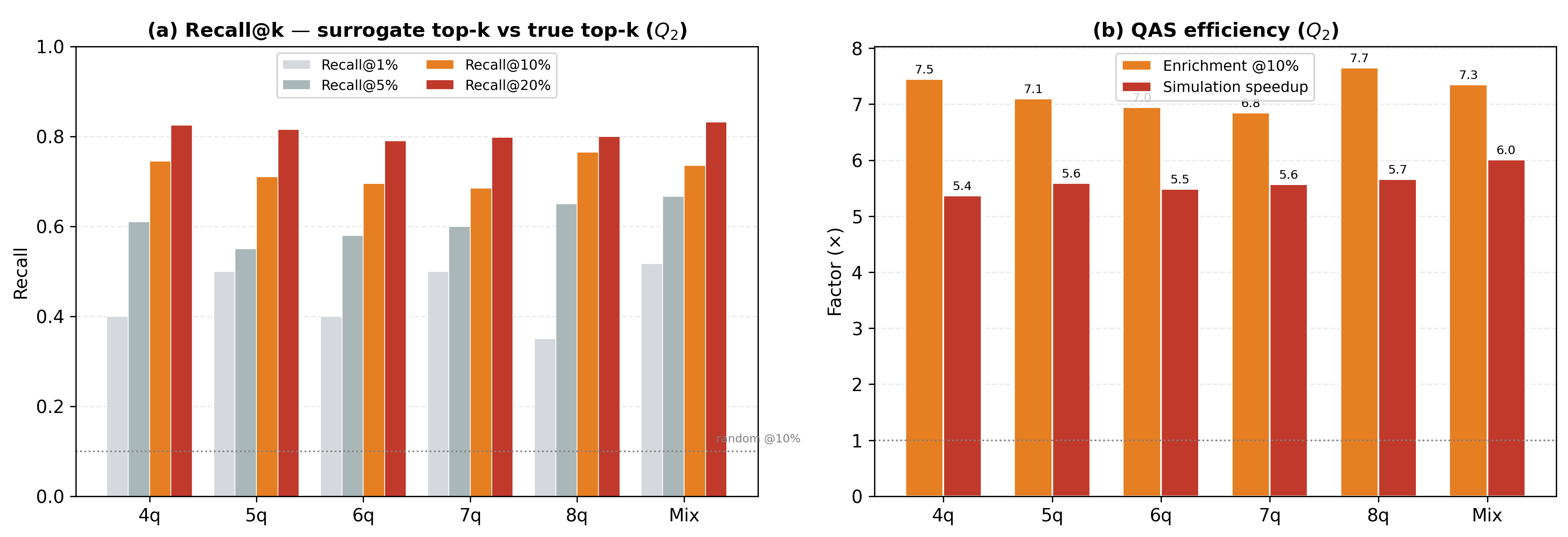}
    \caption{Utility of the surrogate for QAS pre-screening on the held-out test set ($Q_2$; median-seed model). (a) Recall@$k$ between the surrogate's top-$k$ and the true top-$k$, for $k$ at the top $1\%$, $5\%$, $10\%$, and $20\%$. (b) Enrichment factor at the top $10\%$ (concentration of truly top-ranked circuits relative to random selection) and the reduction in statevector simulations required to recover $90\%$ of the true top-$10\%$ circuits, per system size.}
    \label{fig:qasdemo}
\end{figure}

Figure~\ref{fig:scott_corr}  repeats the evaluation for the second-order Scott measure $Q_2$. Over ten seeds the mean RMSE is $0.0391 \pm 0.0006$, $0.0416 \pm 0.0009$, and $0.0396 \pm 0.0008$ for the 4-, 5-, and 6-qubit datasets, $0.0396 \pm 0.0009$ and $0.0389 \pm 0.0010$ for 7 and 8 qubits, and $0.0368 \pm 0.0007$ on the combined dataset, with Spearman $\rho \approx 0.93$--$0.95$ and Kendall $\tau \approx 0.78$--$0.81$-, comparable to the $Q_1$ results. That a higher-order measure is predicted with accuracy similar to the first-order one indicates the model captures circuit structure relevant to both, rather than only single-qubit statistics. As with $Q_1$, predictions in the highest-entanglement region fall slightly below the diagonal, a mild under-prediction consistent with the reduced top-percentile recall in Figure~\ref{fig:qasdemo}.
\begin{figure}[!t]
    \centering
    \includegraphics[width=\columnwidth]{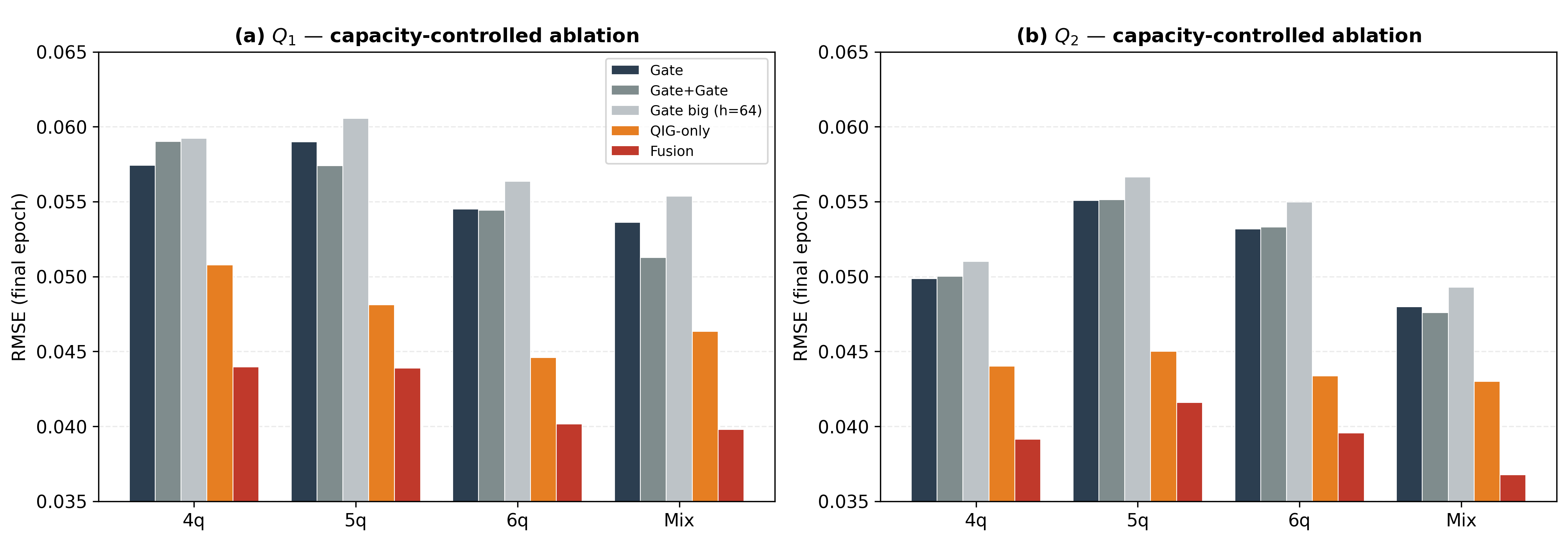}
    \caption{Final-epoch RMSE (mean over 10 seeds) for five encoder variants on the 4-, 5-, 6-qubit datasets and their combined set (Mix), for $Q_1$ (a) and $Q_2$ (b). Variants: gate-level baseline (278k parameters), a parameter-matched dual gate--gate encoder (555k), a widened single gate encoder ($h=64$, 570k), the QIG-only encoder (277k), and the proposed QIG-Fusion (554k). The dashed line marks the baseline level.}
    \label{fig:fiveencoder}
\end{figure}
To determine whether the fusion improvement reflects the qubit-interaction information or merely its larger parameter count, we compared it against two parameter-matched controls (Figure~\ref{fig:fiveencoder}). A dual gate--gate encoder, which has the same two-branch architecture and parameter count as the fusion but receives gate-level input in both branches, performs no better than the single-branch baseline ($Q_2$-Mix: $0.0476$ vs.\ $0.0480$); widening the baseline to an equal parameter budget (gate-big) likewise yields no improvement. In contrast, the QIG-only encoder, with fewer parameters than the baseline, already surpasses it across all cells ($Q_2$-Mix: $0.0430$ vs.\ $0.0480$), and the fusion is best throughout ($0.0368$). A smaller model outperforming a larger one through a different representation cannot be attributed to capacity, isolating the qubit-interaction information as the source of the gain.

Figure~\ref{fig:qasdemo} assesses the surrogate's practical value as a pre-screening stage for QAS. Recall@$k$ increases as the screening fraction relaxes, from $0.35$--$0.52$ at the top $1\%$ to $0.69$--$0.77$ at the top $10\%$ and $0.79$--$0.83$ at the top $20\%$, indicating that the surrogate is most effective for broad candidate reduction rather than for pinpointing the single optimum. Its limited recall at the extreme top percentile is consistent with the mild under-prediction of the highest-entanglement circuits noted earlier. The predicted top $10\%$ is approximately $7\times$ enriched in truly top-ranked circuits relative to random screening, and recovering $90\%$ of the true top-$10\%$ requires about $5.4$--$6.0\times$ fewer statevector simulations than random search-, -a concrete reduction in the number of expensive ground-truth evaluations a practitioner must perform.

\begin{figure}[!t]
    \centering
    \includegraphics[width=\columnwidth]{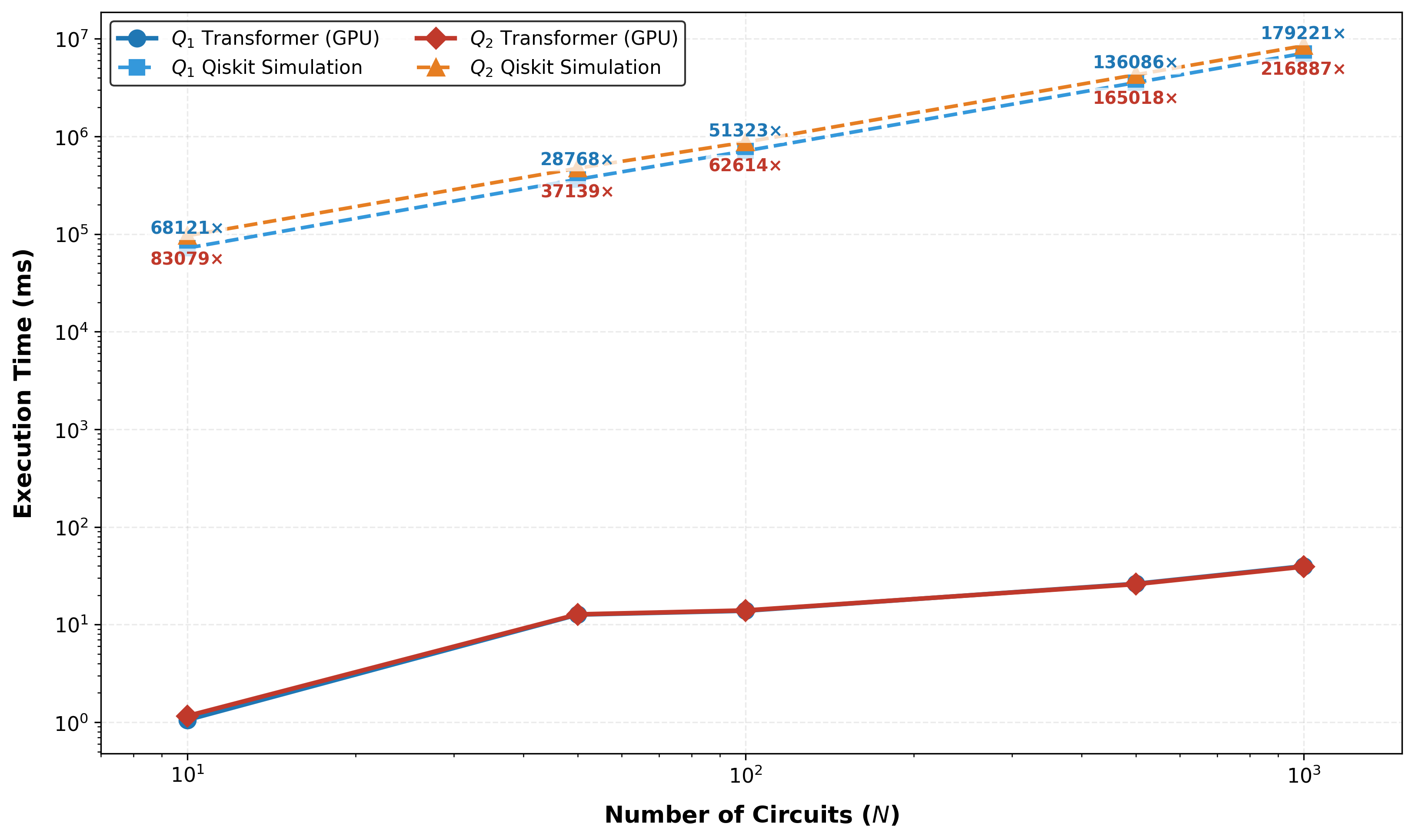}
    \caption{Wall-clock time to evaluate batches of 6-qubit circuits with the trained surrogate (single GPU, batched forward pass) versus the ground-truth protocol (Qiskit statevector estimation, $5{,}000$ parameter samples per circuit), as a function of batch size. Annotations give the surrogate's speed-up over the ground truth at each batch size.}
    \label{fig:inferencespeed}
\end{figure}

Figure~\ref{fig:inferencespeed} reports the per-circuit cost of surrogate inference against the sampling-based ground truth. Because a single forward pass amortizes across a batch while the ground-truth protocol scales linearly with the number of circuits, the speed-up grows with batch size, from $\approx 6.8\times10^{4}\times$ at 10 circuits to $\approx 1.8\times10^{5}\times$ at $1{,}000$ circuits. This measures the cost of evaluating a candidate circuit and is distinct from the search efficiency of Figure~\ref{fig:qasdemo}: the former reflects raw evaluation speed, while the latter reflects the surrogate's ranking fidelity. We report the two separately, since the $10^{5}\times$ per-evaluation speed-up should not be conflated with the more modest but operationally decisive reduction in the number of expensive simulations ultimately required.

\begin{figure*}[!htbp]
    \centering
    \includegraphics[width=\textwidth]{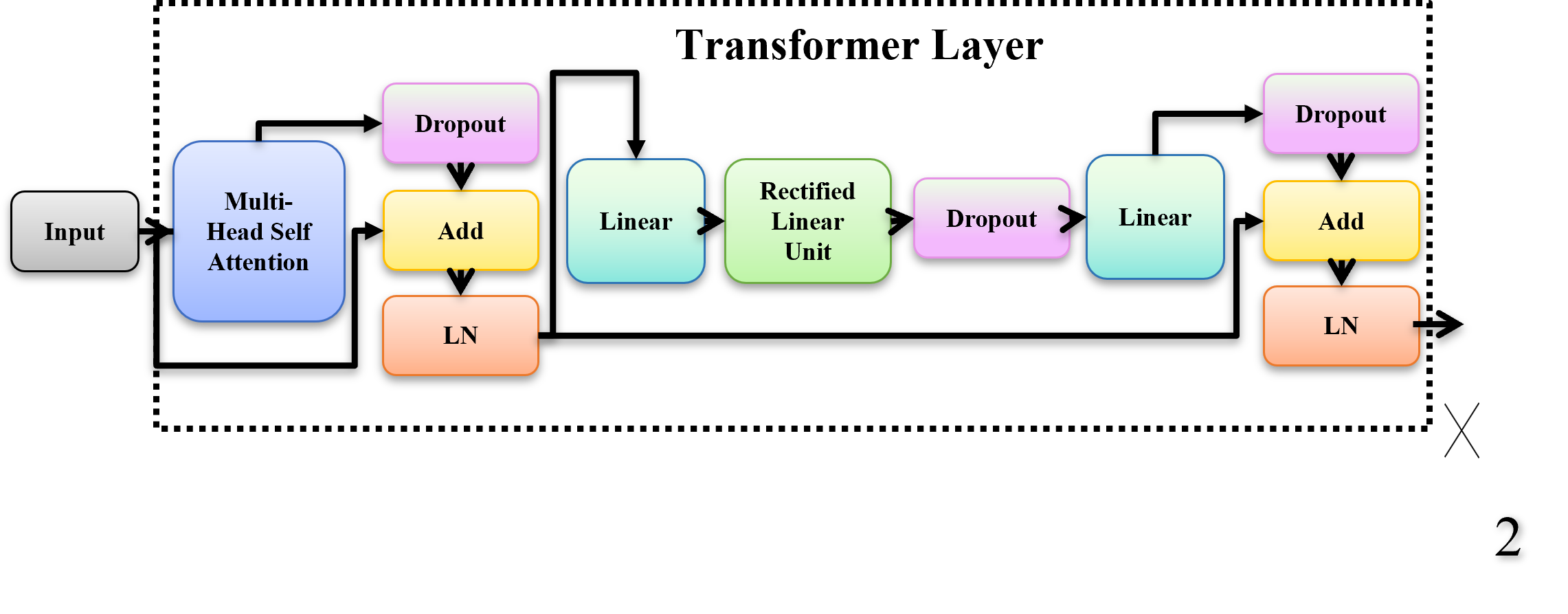}
    \caption{The proposed graph-based transformer architecture for predicting the entangling capability of parameterized quantum circuits. The network processes the directed acyclic graph (DAG) representation and utilizes a sigmoid activation at the output layer to bound the prediction within $[0, 1]$.}
    \label{fig:architecture}
\end{figure*}

Figure~\ref{fig:architecture} illustrates the transformer architecture for predicting PQC entangling capability. This architecture is designed to process graph representations of circuits. The processing pipeline begins with the output of graph embedding module in Figure~\ref{fig:pipeline}. The initial circuit representation, encoded as a DAG consisting of gate types and an adjacency matrix, is projected into a high-dimensional continuous feature vector space through a linear transformation. This initial embedding serves as input to the main processing block.

The core unit consists of a stack of $L$ identical transformer encoder layers operating sequentially. Each encoder layer consists of two main submodules equipped with residual connections and a normalization layer (LN). The first submodule uses a multi-head self-attention (MSA) mechanism to capture long-range dependencies and topological correlations among gates in the circuit, followed by dropout regularization. The second submodule is a position-based feed-forward network (FFN), consisting of two linear layers separated by a ReLU activation function and an additional dropout layer. This structure allows the model to extract complex hierarchical features from the circuit graph topology. After passing through $L$ transformer layers, the enriched feature representation is passed to the output head to produce the final scalar prediction. This module consists of a fully connected (Linear) layer followed by a LeakyReLU activation function to introduce non-linearity, and a second Linear layer to condense the features into a single value. Most importantly, a sigmoid activation function is applied at the final stage to strictly constrain the final Entanglement Output within the range $[0,1]$, 0 is a product state and 1 is a maximally entangled state. This constraint is necessary to ensure that the model output remains consistent with the mathematical definition of the entanglement measure. For QIG-Fusion, the outputs of the two identical transformer encoders are first aggregated using mean pooling. The pooled embeddings are then concatenated before the final prediction layer, following the architecture described in the theory section.

\begin{figure*}[!htbp]
    \centering
    \includegraphics[width=0.85\textwidth]{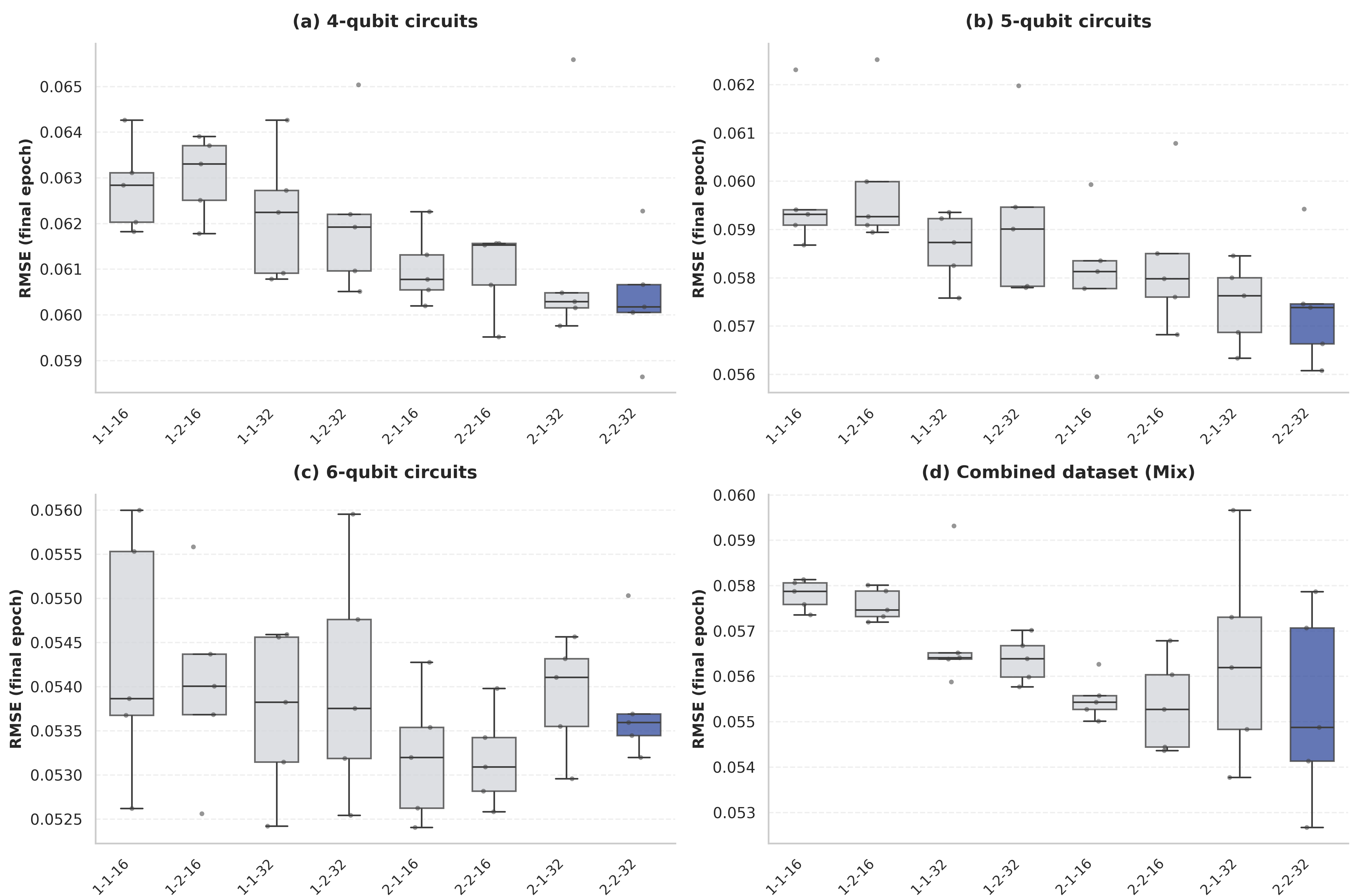}
    \caption{RMSE comparison of $Q_1$ predictions across transformer architectures. The $(2-2-32)$ structure is used for entangling capability prediction.}
    \label{fig:ablation}
\end{figure*}

A comprehensive architecture ablation study evaluating the predictive performance of various model configurations on $Q_1$ was conducted to select the best structure from among the various transformer model structures by selecting the structure that yields significantly varying prediction errors, as illustrated in the Figure~\ref{fig:ablation}. Model performance was evaluated using the RMSE measured at the final training stage (epoch 100). The horizontal axis represents different architectural configurations expressed in triplet format, typically corresponding to parameters such as layer depth, number of attention heads, and embedding dimension ($L - H - D$). To improve statistical reliability and capture stochastic variations during training, all results are averaged across five independent random seeds. Evaluations are performed on separate datasets corresponding to different circuit sizes, as well as a combined dataset containing all configurations. A clear overall pattern is observed across all datasets as increasing model complexity ($L - H - D$) generally results in lower RMSE values at convergence. This indicates that $Q_1$ size predictions benefit from higher model capacity and richer representation power. For smaller circuits (4 and 5 qubits), the baseline configuration $(2 - 2 - 32)$ consistently achieves the best performance, with the lowest median RMSE and relatively narrow variability, indicating its robustness. A slight deviation appears in the 6-qubit case. While the baseline configuration remains competitive, the $(2 - 1 - 16)$ and $(2 - 2 - 16)$ configurations achieve slightly lower median RMSE values. This suggests that for 6-qubit circuits, excessive model capacity can lead to mild overfitting, or alternatively, a lower-dimensional representation is sufficient for generalization. In the mixed dataset, the baseline configuration achieves the lowest absolute RMSE across all seeds and demonstrates robust results. Therefore, the $(2 - 2 - 32)$ structure is used for entangling capability prediction. The same applies to $Q_2$.

Figure~\ref{fig:loss} presents the Mean Squared Error (MSE) convergence curves of the optimal Transformer $(2 - 2 - 32)$ model over 100 epochs for the $Q_1$ and $Q_2$ metrics. These curves represent the average of 10 independent seeds. The shaded area indicates the standard deviation band ($\pm 1\sigma$). Overall, all qubit scenarios (4, 5, 6, and Mix) exhibit excellent convergence profiles, characterized by a smooth exponential decrease in the initial phase before eventually reaching a plateau. The gradient slope over the first 20 epochs demonstrates the efficiency of the MSA mechanism in mapping the DAG structural features of the quantum circuit to its entanglement values. The stability of the test loss, which shows no divergence after epoch 50, provides strong empirical evidence that the proposed architecture is highly robust against overfitting. This is confirmed by the narrow $\pm 1\sigma$ band across the curve, confirming the model's consistent performance regardless of the stochastic weight initialization (10-seed). However, this model also exhibits higher variability compared to simpler architectures, indicating greater sensitivity to initialization. At epoch 100, the model achieves robust performance with a very low final MSE value. For the 4- and 5-qubit cases, the training loss is around 0.0025, while the testing loss is around 0.0035. For the 6-qubit and mixed datasets, the model achieves deeper optimization, with training losses approaching 0.0020 and testing losses around 0.0030. The consistently small difference between training and testing losses (around 0.001) indicates good generalization to previously unseen data.

A small anomaly was observed in the Mix model, where the test loss consistently exceeded the training loss by a larger margin than the single-qubit scenario ($\sim 0.003$ vs $\sim 0.002$ at the end of the epoch). Several possibilities exist, including a potential slight distribution shift, given that the Mix model was trained on 24,000 heterogeneous circuits (a mix of 4/5/6 qubits), which increases the complexity of the feature space compared to the homogeneous model. Furthermore, it is possible that the nested partitioning in the combined dataset does not fully proportionally represent the variation in the number of gates between the training and test sets. Despite the observed anomaly, overall prediction performance remained unaffected. In fact, the mixed dataset model achieved the best RMSE value in the final evaluation. These results indicate that, even with slight inconsistencies in the loss values, the model maintains strong generalization ability and ranking accuracy, as supported by the high Spearman correlation and Kendall's Tau.

\begin{figure*}[!htbp]
    \centering
    \includegraphics[width=0.9\textwidth]{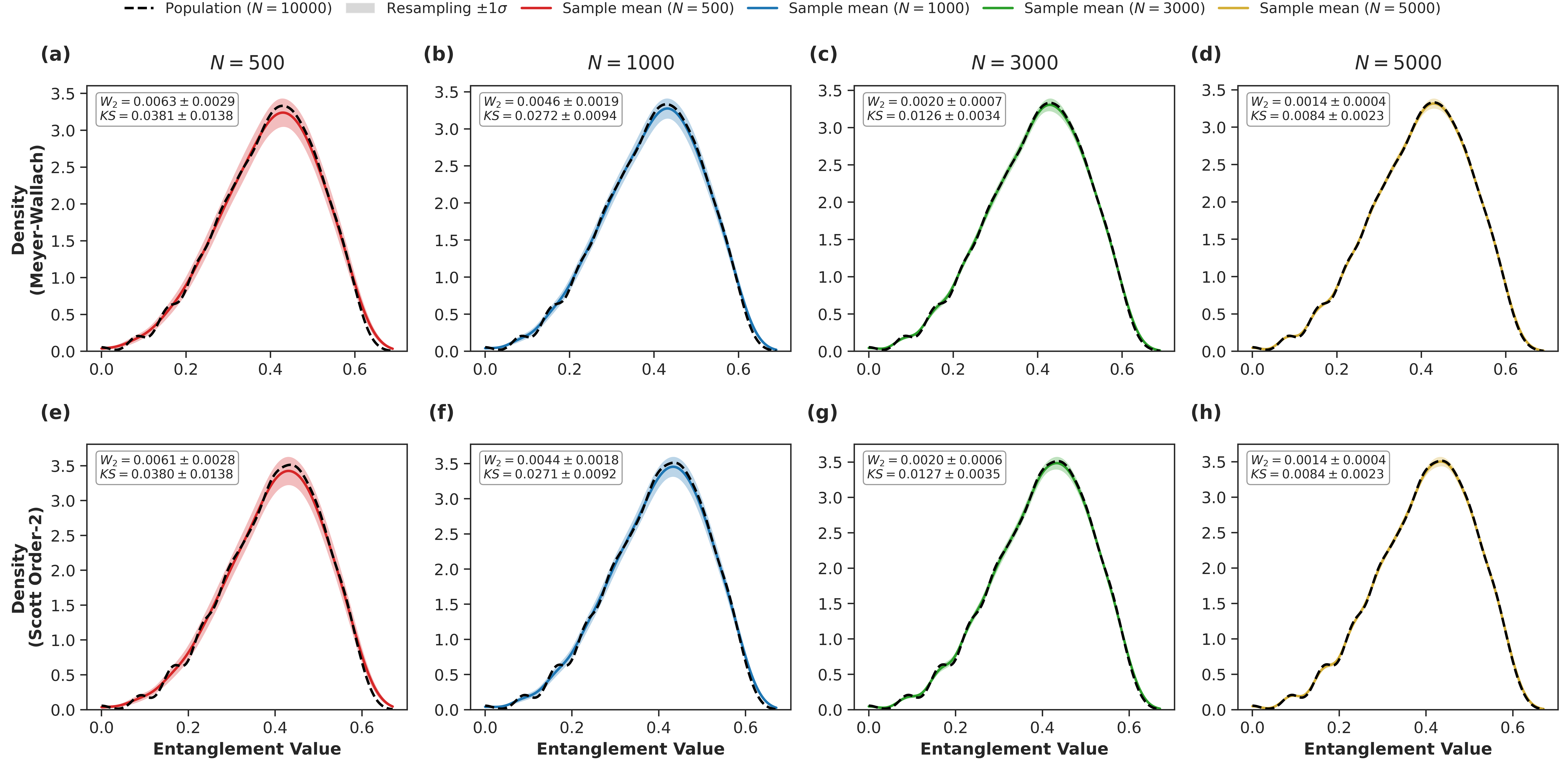}
    \caption{Distribution stability under resampling for $Q_1$ \& $Q_2$ (Qubit-6, 100 resamples per panel), as the sampling size increases from 500 to 5000, a clear convergence in the distributions can be observed, shows that $N = 5000$ samples are sufficient.}
    \label{fig:resampling}
\end{figure*}

To ensure sufficient diversity in the resulting PQCs, a kernel density estimation analysis~\cite{Parzen1962, Rosenblatt1956} was performed on entanglement values across various sample sizes, as illustrated in Figure~\ref{fig:resampling}. The entanglement density distribution for the 5,000-circuit sample empirically replicates the shape of the 10,000-circuit population curve (represented by the solid black line). This indicates that sampling 10,000 circuits per qubit size produces a highly stable and statistically convergent distribution, which is more than sufficient to adequately train the transformer model.

\begin{figure*}[!htbp]
    \centering
    \includegraphics[width=0.85\textwidth]{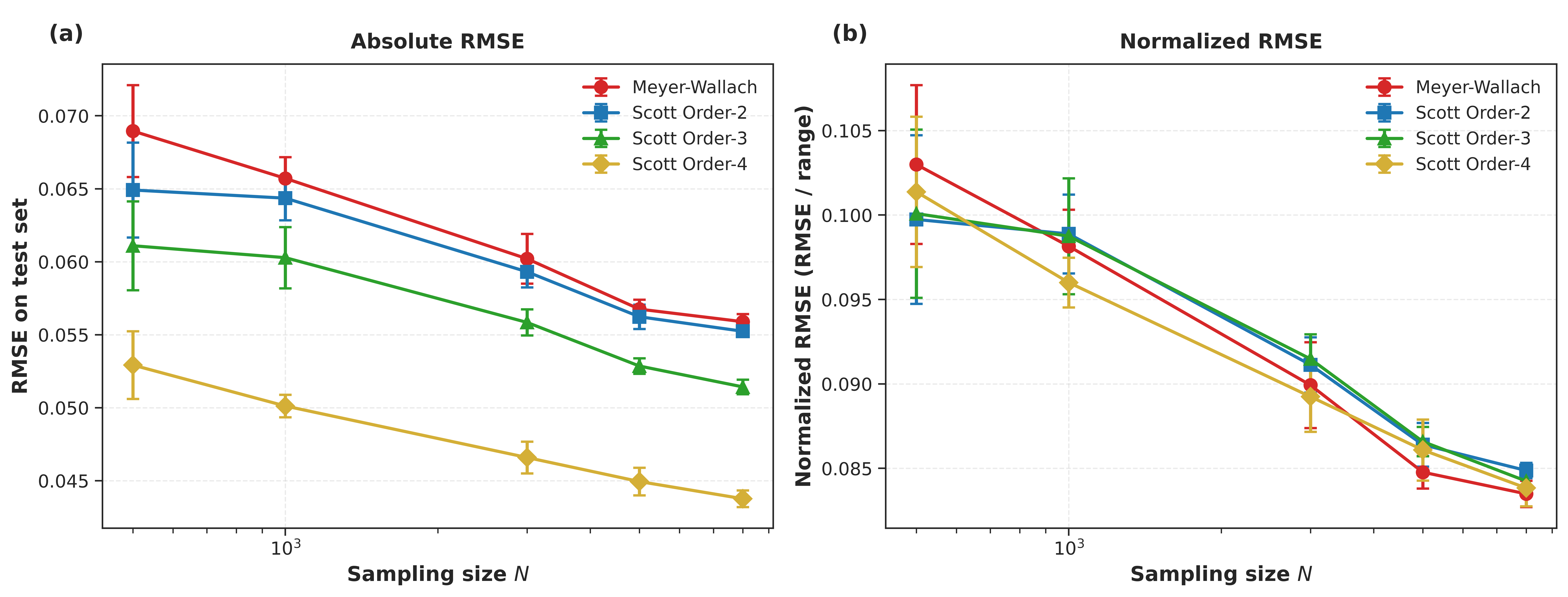}
    \caption{RMSE Convergence with Sampling Size (Qubit-6) Normalized by value range per metric.}
    \label{fig:rmse_size}
\end{figure*}

Figure~\ref{fig:rmse_size} presents the RMSE as a function of the training set size $N$, evaluated on a fixed test set of 2,000 circuits. Panel (a) shows that the absolute RMSE decreases monotonically for all four entanglement sizes. However, because different sizes exhibit different value ranges, we also report the normalized RMSE (RMSE divided by the value range) in panel (b). Normalized results show that all four measures converge to comparable prediction difficulty at $N = 8000$, the rate of improvement decreases significantly beyond $N = 5000$ (marginal improvement $< 1.5\%$), and Scott Order-4 requires more training samples to stabilize, consistent with its more complex multipartite structure. These results justify our choice of 10,000 circuits per qubit calculation as an adequate dataset size.

\begin{figure*}[!htbp]
    \centering
    \includegraphics[width=0.9\textwidth]{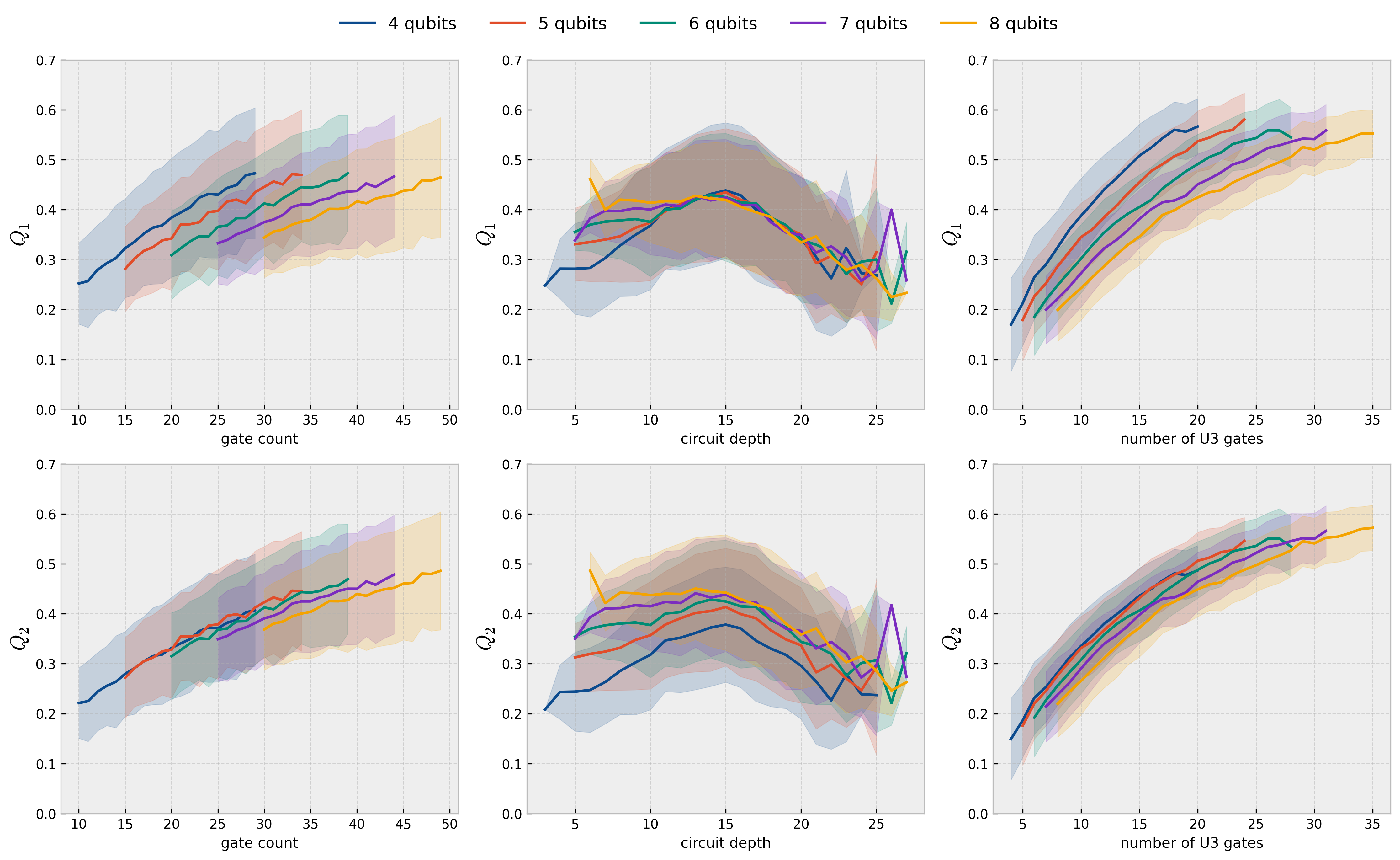}
    \caption{Correlation between $Q_1$, $Q_2$ and quantum circuit characteristics (gate count, circuit depth, and $U_3$ gates).}
    \label{fig:characteristics}
\end{figure*}

This dataset was generated with a gatewise pipeline with softmax-based gate selection drawn from a Gaussian distribution which naturally limits the $CZ$ gate ratio to approximately $20 - 85\%$ of the total gate count and enforces a linear nearest-neighbor connectivity topology, consistent with the limitations of superconducting hardware such as the OriginQ Wukong device~\cite{Kong2021}; As a result, the resulting entangling capability values $Q_1$ and $Q_2$ span the range approximately $[0.00, 0.67]$, reflecting the physical limitations of the circuit family rather than sampling artifacts, since circuits with zero $CZ$ gates, which yield $Q_1 = 0$ deterministically, are structurally excluded by the opening $U_3$ layer requirement, and circuits with highly regular alternating $CZ - U_3$ patterns that can approach the theoretical limit of Haar randomness in Equation~(\ref{eq:haar}) is statistically underrepresented under random softmax sampling. Figure~\ref{fig:characteristics} reveals a consistent and physically meaningful relationship between the circuit's structural properties. For 4-qubit circuits, the maximum achievable values gradually decrease: $Q_1 \approx 0.60$ and $Q_2 \approx 0.55$. This hierarchy is not an artifact of the model, but rather a direct manifestation of the geometric constraints described in the theory section, approaching the Haar limit only when $n - m \gtrsim m$. A clear positive correlation is observed between the entanglement value and the total number of gates and the number of parameterized rotation gates ($U_3$). Increasing the number of quantum operations leads to a higher ability of the circuit to generate more complex quantum correlations. From a physical perspective, this behavior is intuitive, as additional unitary transformations enlarge the accessible Hilbert space and increase the system's degrees of freedom. Consequently, the circuit can explore more complex multipartite entangled states. The number of parameterized rotation gates exhibits smooth, monotonic, and nearly saturating behavior across $Q_1$ and $Q_2$ without bell-shaped artifacts. This is physically expected: the $U_3$ gate is the only free parameter carrier in our gate set; a $CZ$ gate without prior rotation cannot generate entanglement. Consequently, for Quantum Architecture Search applications, the $U_3$ density emerges as a more reliable a priori indicator of entangling capability than the total gate number or nominal circuit depth. We anticipate this insight will inform the design of variational analyses targeting specific entanglement profiles.

A similar monotonic trend is observed with respect to the number of parameterized gates, where a higher tunable rotation density allows for more flexible state preparation. This reinforces the idea that parameterization plays a crucial role in shaping the entanglement landscape, as it directly controls the expressivity and adaptability of the circuit during sampling. However, a very different behavior emerges when analyzing circuit depth. Both metrics exhibit a non-monotonic dependence on circuit depth $d$, with entanglement increasing sharply for $d \in [4, 14]$, peaking near $d \approx 15$, and then decreasing for $d > 17$.

This trend is governed by gate composition rather than by depth itself. Across the 4/5/6-qubit ensembles (30,000 circuits), the $U_3$-gate ratio is the dominant and consistent predictor of
entanglement (Pearson $r = 0.51, 0.60, 0.64$ for 4, 5, 6 qubits, increasing with system size), whereas raw circuit depth is only weakly correlated and weakens with size ($r =
0.26, 0.11,\sim 0.00$). Because the gatewise generator holds the gate budget loosely fixed, depth and $U_3$ ratio are strongly anticorrelated ($r \sim - 0.46 \text{ to} - 0.52$): deeper circuits allocate a larger fraction of gates to CZ and contain proportionally fewer $U_3$ gates. The apparent decline of $Q_1$ and $Q_2$ beyond the mid-depth maximum therefore tracks the falling $U_3$ ratio, and the high-depth tail ($d > 22$) is additionally sparsely sampled, where each bin contains only a few circuits and the means become noisy.

These circuits do not approach the Haar (t-design) regime. On the 6-qubit ensemble, the second Renyi entropy on a balanced 3-3 bipartition reaches at most $\sim 17 \%$ of the Haar (Page) value (maximum ratio 0.167), and the frame potential ratio $\frac{F^2}{F_{\text{Haar}}^2}$ remains at least 1.73 at all depths
(an exact 2-design would give 1.0) without decreasing toward unity as depth grows. This is the expected behaviour for a one-dimensional nearest-neighbour architecture, for which approximate t-design formation requires depth of order $\mathcal{O}(n* poly(t))$ \cite{Brandao2016,Harrow2023}, well beyond the depth range accessible here ($d \leq 27$). We therefore interpret the depth trend as a consequence of parameterized-gate density rather than t-design formation, consistent with our finding that $U_3$ density is the strongest a-priori indicator of entangling capability.
\section{Conclusion}
We developed a graph-based transformer framework for predicting the multipartite entanglement structure of PQCs through the Meyer--Wallach measure ($Q_1$) and the second-order Scott measure ($Q_2$). To our knowledge, $Q_2$ has not previously been targeted by a learned circuit surrogate; predicting it alongside $Q_1$ provides a richer characterization of entangling capability. Our central methodological contribution is the qubit-interconnected graph (QIG) encoding, in which each node represents a qubit and a weighted adjacency records entangler multiplicity on each bond; fused with a gate-level DAG encoder that preserves gate parallelism, circuit depth, and temporal dependency, this yields the QIG-Fusion model.\\
\indent The proposed QIG-Fusion achieved a consistent improvement RMSE over the gate-level baseline for both $Q_1$ and $Q_2$ measures. Capacity-matched ablations confirmed that this gain stems from the qubit-interaction information rather than from added parameters. A parameter-matched dual gate-level encoder conferred no benefit, while a qubit-graph encoder with fewer parameters already surpassed the baseline. Because the surrogate is optimized to preserve the ranking of candidate circuits rather than to reproduce exact values, its objective aligns with the practical requirements of Quantum Architecture Search (QAS).\\
\indent Across a ten-seed protocol, QIG-Fusion achieved RMSE of $0.037$--$0.042$ ($Q_2$) and $0.040$--$0.044$ ($Q_1$), a consistent improvement over the gate-level baseline ($0.048$--$0.059$), with Spearman rank correlation up to $0.95$. Capacity-matched ablations confirmed that this gain stems from the qubit-interaction information rather than from added parameters: a parameter-matched dual gate-level encoder conferred no benefit, while a qubit-graph encoder with fewer parameters already surpassed the baseline. Because the surrogate is optimized to preserve the ranking of candidate circuits rather than to reproduce exact values, its objective aligns with the practical requirements of Quantum Architecture Search (QAS). A screening-utility demonstration confirmed this alignment: the surrogate concentrated high-entanglement circuits with an enrichment factor of $\approx 7\times$ over random selection and reduced the number of statevector simulations required to recover $90\%$ of the true top-$10\%$ circuits by $\approx 5.4$--$6.0\times$; this search-level efficiency is complementary to the per-circuit inference speedup of $\approx 6.8\times10^{4}$--$1.8\times10^{5}\times$ over the ground-truth protocol. Finally, we probed scalability by extending to larger Hilbert spaces ($7$--$8$ qubits), where circuit size grows in qubit count ($4\!\to\!8$), gate count ($\approx\!12\!\to\!32$ on average), and depth ($\approx\!11.6\!\to\!13.9$ on average) simultaneously; the QIG-Fusion advantage persisted and, in rank correlation, grew with system size. Beyond predictive performance, our results highlight physical aspects of entangling capability in PQCs. Both $Q_1$ and $Q_2$ increase with gate count and circuit depth while the gate composition ($U_3$ fraction $\approx 0.55$) remains approximately constant across sizes (Figure~\ref{fig:characteristics}). Within the studied regime, the measures exhibit diminishing returns beyond a moderate depth of roughly $15$ layers, showing little further increase for deeper circuits. This indicates that highly entangled states are attainable without excessively deep circuits, a useful guideline for circuit design in QAS, where depth directly determines computational and noise cost.\\
\indent Our framework opens several research avenues. First, it can be extended to additional descriptors such as expressibility, or entangling capability jointly with trainability metrics (e.g., gradient variance for barren-plateau detection) within a single graph-attention pipeline. We note, however, that a meaningful trainability study requires an ensemble that reaches the barren-plateau regime; our present random-circuit dataset lies far from a unitary 2-design, so this extension calls for a purpose-built dataset. A second direction is to enrich the qubit-interconnected graph beyond the linear-nearest-neighbour topology studied here, toward richer connectivities where the interaction structure is denser and the QIG advantage is expected to be more pronounced. The architecture is size-independent and scales favourably with circuit size. The gate-level encoding uses a feature matrix of size $N_g \times (N_{gt}+N_q)$---where $N_g$ is the number of gates, $N_{gt}$ the number of gate categories, and $N_q$ the number of qubits---and an adjacency matrix of size $N_g \times N_g$; the qubit-interconnected graph adds only an $N_q \times N_q$ adjacency matrix and an $N_q \times (2+N_q)$ feature matrix. The overall cost is therefore dominated by the $\mathcal{O}(N_g^2)$ gate adjacency and grows only linearly in the number of gate types and qubits, avoiding the exponential scaling of direct simulation. Given a labeled dataset at a larger qubit count, the model can thus estimate entangling capability without a substantial increase in overhead; transfer learning from smaller circuits, a direction also noted by Zhang et al, is a natural way to amortize the labeling cost. Finally, a key application is to close the loop with downstream variational tasks. Circuits identified as highly entangling by our surrogate could serve as structured ansatz candidates for variational quantum algorithms, for instance, as initial ansätze in the variational quantum eigensolver for molecular ground-state energy estimation---allowing the practical value of entanglement-guided circuit selection to be assessed directly on physical objectives. Integrating the surrogate as a pre-screening stage within a full closed-loop QAS search, and evaluating whether entanglement-guided selection accelerates convergence on such tasks, is a promising direction toward entanglement-aware variational quantum algorithms in the NISQ era.

\section*{Data Availability Statement}

The data that support the findings of this study are openly available in GitHub at \url{https://github.com/DarellTim/predict_multipartite_entanglement_transformer}.

\begin{acknowledgments}
This work was supported by grants from National Research and Innovation Agency (BRIN), Indonesia and Centre for Quantum Technologies (CQT), National University of Singapore, Singapore.
\end{acknowledgments}

\end{document}